\documentclass[aps,nofootinbib,showpacs,twocolumn,superscriptaddress,10pt]{revtex4-1}

\usepackage{natbib}
\usepackage{amsmath,amssymb}
\usepackage{graphics}

\newcommand{\ket}[1]{|{#1}\rangle}

\newcommand{\braket}[2]{\langle{#1}|{#2}\rangle}

\begin{document}

\title{Experimentally observed decay of higher-dimensional \\ entanglement through turbulence}

\author{Yingwen \surname{Zhang}}
\affiliation{CSIR National Laser Centre, PO Box 395, Pretoria 0001, South Africa}

\author{Shashi \surname{Prabhakar}}
\affiliation{CSIR National Laser Centre, PO Box 395, Pretoria 0001, South Africa}
\affiliation{School of Physics, University of the Witwatersrand, Johannesburg 2000, South Africa}

\author{Alpha \surname{Hamadou Ibrahim}}
\affiliation{CSIR National Laser Centre, PO Box 395, Pretoria 0001, South Africa}

\author{Filippus S. \surname{Roux}}
\email{fsroux@csir.co.za}
\affiliation{CSIR National Laser Centre, PO Box 395, Pretoria 0001, South Africa}
\affiliation{School of Physics, University of the Witwatersrand, Johannesburg 2000, South Africa}

\author{Andrew \surname{Forbes}}
\affiliation{School of Physics, University of the Witwatersrand, Johannesburg 2000, South Africa}

\author{Thomas \surname{Konrad}}
\affiliation{University of Kwazulu-Natal, Private Bag X54001, Durban 4000, South Africa}
\affiliation{National Institute of Theoretical Physics, Durban Node, South Africa}

\begin{abstract}
The evolution of high dimensional entanglement in atmospheric turbulence is investigated. We study the effects of turbulence on photonic states generated by spontaneous parametric down-conversion, both theoretically and experimentally. One of the photons propagates through turbulence, while the other is left undisturbed. The atmospheric turbulence is simulated by a single phase screen based on the Kolmogorov theory of turbulence. The output after turbulence is projected into a three-dimensional (qutrit) basis composed of specific Laguerre-Gaussian modes. A full state tomography is performed to determine the density matrix for each output quantum state. These density matrices are used to determine the amount of entanglement, quantified in terms of the negativity, as a function of the scintillation strength. Theoretically, the entanglement is calculated using a single phase screen approximation. We obtain good agreement between theory and experiment.
\end{abstract}

\pacs{03.67.Hk, 03.65.Yz, 42.50.Tx, 42.68.Bz}

\maketitle

\section{Introduction}

There are advantages in the use of high dimensional entangled systems in quantum information processing and communication. Such systems with dimensions larger than 2 allow for a higher information capacity \cite{walborn} and lead to increased security in quantum cryptography \cite{bp2000}. Quantum entanglement is an important resource in many quantum information protocols. However it is fragile, since it decays through interaction with noisy environments \cite{konrad,tiersch}, such as atmospheric turbulence.

The orbital angular momentum (OAM) states of photons are a suitable candidate for the implementation of high dimensional quantum systems for use in high dimensional quantum key distribution \cite{zeil2006,mafu} and long-range quantum communication \cite{capraro} through free-space. Unlike the polarization of light, which offers a two-level Hilbert space, the OAM of photons provides an infinite-dimensional Hilbert space. Fortuitously, photonic states produced in spontaneous parametric down-conversion (SPDC) are naturally entangled in their OAM degrees of freedom due to the conservation of OAM \cite{arnaut,franke,mair}.

The decay of OAM entanglement in a qubit pair evolving in turbulence has been studied both theoretically \cite{sr,qturb4,qturb3,ipe, toddbrun,leonhard} and experimentally \cite{pors,malik,oamturb,qkdturb}. These studies show that qubit OAM entanglement decays slower in weak scintillation for modes with higher OAM-values. While, the study of turbulence induced decay of OAM entanglement in qubit pairs can help one to understand basic behavior of OAM entanglement in turbulence, it offers little practical benefit over the use of the polarization states of single photons, which are less affected by turbulence. The point of using the OAM states of light is to have high dimensional quantum states. Therefore, it makes sense that one should consider high dimensional states in the OAM basis for these studies.

The effect of atmospheric turbulence on such high dimensional OAM entanglement have received little attention. The only case where the evolution of OAM entanglement for high dimensional quantum state has been considered, is a theoretical investigation involving qutrits \cite{bruenner}. It appears that the effects of turbulence on the OAM entanglement for high dimensional quantum states, have not yet been considered experimentally.  

For this reason, we report here on the theoretical and experimental results of a study into the effects of atmospheric turbulence on high dimensional OAM entanglement. Using SPDC, we prepare photon pairs that are entangled in their OAM degrees of freedom. One of these photons is propagated through turbulence, while allowing the other to propagate undisturbed through free-space (without turbulence). The turbulence is implemented as a phase-only distortion on a single phase screen, using a spatial light modulator (SLM). By implication, we use the single phase screen (SPS) approximation of the atmospheric scintillation process \cite{paterson}, which is only valid under weak scintillation conditions \cite{turbsim}. Inspite of this restriction, the SPS model is still currently the most widely used model for the study of photonic quantum states propagating in turbulence \cite{sr,qturb4,qturb3, toddbrun,leonhard,pors,malik,oamturb}. Our results show a good agreement between theory and experiment.

There are various aspects that pose challenges to the experimental study of the propagation of high dimensional OAM entangled states through turbulence. To determine the high dimensional quantum state, one needs to perform a quantum state tomography, which requires a number of individual measurements that increases rapidly with the increasing number of dimensions \cite{tomo}. Moreover, due to the randomness of atmospheric turbulence, one has to repeat these measurements a reasonable number of times and average the results to obtain a meaningful statistical description of the evolution of the quantum state. In other words, an investigation of the decay of high dimensional entanglement in turbulence is a very time consuming process. In our experiment, we consider qutrits --- three-dimensional states --- and performed a full state tomography for each output state obtained from different realizations and different strengths of the turbulence. While three-dimensional states allow us to observe the effects of high dimensions, we are able to perform the experiments over a reasonable period.

An important purpose of this work is to show that we are able to obtain a good agreement between the theoretical predictions that one can compute using the SPS approximation. Here, we use the Laguerre-Gaussian (LG) basis as the measurement basis, because it enables us to obtain relatively simple analytical expressions. For accurate comparison, the experimental measurements need to implement precisely that which is assumed in the calculations. As a result we measure in the exact LG basis and not in the helical basis where the amplitude variation of the mode is ignored, apart from the Gaussian envelop that is imposed by the overlap in the optical fibre. For this reason we use complex amplitude modulation \cite{arrizon1} on the SLMs to perform the projective measurements required for the quantum state tomography.

The paper is organized as follows. Various theoretical aspects are discussed in Sec.~\ref{agter}. The experimental setup is discussed in Sec.~\ref{opstel}, followed by a discussion of the results in Sec.~\ref{resdis}. Conclusions are provided in Sec.~\ref{concl}.

\section{Theory}
\label{agter}

\subsection{SPS model}

In the experiment, we pass one of the photons through turbulence that is simulated by a random phase modulation using an SLM. Hence, the appropriate theory with which these experimental results are to be compared is the SPS approximation. For the biphoton case with only one photon propagating through turbulence, the elements of the output density matrix in the SPS approximation are given by \cite{paterson,lindb}
\begin{eqnarray}
\rho_{mnpq} & = & \int E_m^*({\bf x}_1) E_p^*({\bf x}_2) E_n({\bf x}_3) E_q({\bf x}_4) \nonumber \\
& & \times \psi({\bf x}_1,{\bf x}_2) \psi^*({\bf x}_3,{\bf x}_4) \exp\left[-\frac{1}{2} D_{\theta} (\Delta x) \right] \nonumber \\
& & \times {\rm d}^2 x_1\ {\rm d}^2 x_2\ {\rm d}^2 x_3\ {\rm d}^2 x_4 ,
\label{spsint}
\end{eqnarray}
where $\Delta x= |{\bf x}_1-{\bf x}_3|$ and the input field $\psi({\bf x}_1,{\bf x}_2)$ is the biphoton state obtained from the SPDC process (simply called the SPDC state). The output modes, in terms of which the density matrix is defined, are given by $E_n({\bf x}) = \braket{{\bf x}}{n}$, where $\ket{n}$ represents the chosen basis for the output density matrix. The turbulence is represented by the phase structure function $D_{\theta}(\cdot)$. The photon in the $A$-system ($B$-system) is associated with the coordinate vectors ${\bf x}_1$ and ${\bf x}_3$ (${\bf x}_2$ and ${\bf x}_4$) and with the indices $m$ and $n$ ($p$ and $q$).

\subsection{Structure function}

In the Kolmogorov theory \cite{scintbook}, the phase structure function can be expressed as
\begin{equation}
D_{\theta}(x) = 6.88 \left(\frac{x}{r_0}\right)^{5/3} ,
\label{strukt}
\end{equation}
in terms of the Fried parameter \cite{fried},
\begin{equation}
r_0 = 0.185 \left(\frac{\lambda^2}{C_n^2 z}\right)^{3/5} .
\label{fried}
\end{equation}
Here $C_n^2$ is the structure constant and $\lambda$ is the wavelength of the down-converted photons. For degenerate SPDC, $\lambda=2\lambda_{\rm p}$, where $\lambda_{\rm p}$ is the pump wavelength.

Due to the power of $5/3$ in Eq.~(\ref{strukt}), the integral in Eq.~(\ref{spsint}) is not easy to evaluate. Therefore, we use the quadratic structure function approximation \cite{leader}, which implies that one can replace $x^{5/3}\rightarrow x^2$. The resulting quadratic structure function then takes the form 
\begin{equation}
D'_{\theta}(x) = 6.88 \left(\frac{x^2}{w_{\rm p}^{1/3}r_0^{5/3}}\right) .
\label{struktq}
\end{equation}
The extra factor of $w_{\rm p}^{-1/3}$ is necessary to retain a dimensionless argument in the exponential function in Eq.~(\ref{spsint}). Here $w_{\rm p}$ is the radius of the pump beam waist.

\subsection{Input (SPDC) state}

The input state in the SPS calculation is the SPDC state. For the purpose of the theoretical calculation, one can represent the SPDC state as the product of the pump profile in the Fourier domain and the phase matching function that governs the SPDC process in the nonlinear crystal. The pump is assumed to be a Gaussian beam.

The details of the phase matching function is not important, because, as it turns out, the experimental conditions are such that the phase matching function plays a diminished role. (See the discussion of the thin-crystal limit below.) For that reason we follow \cite{eberly} and use the (analytically more tractable) Gaussian function model to represent the phase matching function. 

The resulting SPDC input state in the Fourier domain can therefore be represented by 
\begin{eqnarray}
\Psi'_{\rm spdc}({\bf a}_{\rm s},{\bf a}_{\rm i}) & = & \braket{{\bf a}_{\rm s},{\bf a}_{\rm i}}{\Psi_{\rm spdc}} \nonumber \\
& = & {\cal P'} \exp \left(-\pi^2 w_{\rm p}^2 |{\bf a}_{\rm s}+{\bf a}_{\rm i}|^2\right) \nonumber \\
& & \times \exp \left(-\frac{1}{2} \pi^2 w_{\rm p}^2 \beta |{\bf a}_{\rm s}-{\bf a}_{\rm i}|^2 \right) ,
\label{spdcin}
\end{eqnarray}
where ${\cal P'}$ is a normalization constant, ${\bf a}$ is a two-dimensional spatial frequency vector (related to the transverse propagation vector by
${\bf k}_{\perp} = 2\pi{\bf a}$), and the subscripts $s$ and $i$ denote the two (`signal' and `idler') down-converted photons, respectively. Furthermore, we defined the dimensionless parameter
\begin{equation}
\beta = \frac{n_{\rm o}L\lambda_{\rm p}}{\pi w_{\rm p}^2} = \frac{n_{\rm o}L}{z_{Rp}} ,
\label{beta}
\end{equation}
where $L$ and $n_{\rm o}$ are the length and the ordinary refractive index of the nonlinear crystal, respectively, and $z_{Rp}$ is the Rayleigh range of the pump beam ($\pi w_{\rm p}^2/\lambda_{\rm p}$).

In most experiments, the Rayleigh range of the pump beam is several orders of magnitude larger than the thickness of the nonlinear crystal, $L\ll z_{Rp}$. This leads to the so-called {\em thin crystal limit}, where the phase matching function is only evaluated at the origin. It implies that one can set $\beta=0$ and drop the phase matching function in the calculations. However, the phase matching function may help to regularize the integrals and can be removed at the end by taking the limit $\beta\rightarrow 0$. 

The inverse Fourier transform of the SPDC state, given in Eq.~(\ref{spdcin}), is 
\begin{eqnarray}
\Psi_{\rm spdc}({\bf x}_{\rm s},{\bf x}_{\rm i}) & = & {\cal F}^{-1} \left\{ \Psi'_{\rm spdc}({\bf a}_{\rm s},{\bf a}_{\rm i}) \right\} \nonumber \\
& = & {\cal P} \exp \left(-\frac{|{\bf x}_{\rm s}+{\bf x}_{\rm i}|^2}{4w_{\rm p}^2} -\frac{|{\bf x}_{\rm s}-{\bf x}_{\rm i}|^2}{2 w_{\rm p}^2\beta} \right) . \nonumber \\
\label{spdcinx}
\end{eqnarray}
The normalization constant ${\cal P}$ contains ${\cal P}'$ and includes addition dimension parameters. These normalization constants will eventually drop out of the expression, when the projected state is renormalized.

\subsection{Effective pump width}

The coincidence counts that are produced in the experiment can be predicted in the thin-crystal limit by the follow three-way overlap integral
\begin{equation}
C_{\ell} \propto \int m_p({\bf x}) m_s^*({\bf x}) m_i^*({\bf x})\ {\rm d}^2x ,
\label{oorvl}
\end{equation}
where $m_{p,s,i}({\bf x})$ represent the mode profiles of the pump, signal and idler beams.

The pump is a Gaussian beam, with a particular beam radius $w_{\rm p}'$. However, one needs to take the Gaussian overlap functions coming from the coupling of the light into the SMFs, into account in the overlap calculation. Due to the way in which the measurement basis is encoded on the SLMs, one cannot incorporate the Gaussians from the SMFs into the measurement mode profiles. Therefore, we combine them with the pump profile to produce an effective (Gaussian) pump profile, with an effective pump mode size, given by
\begin{equation}
\frac{1}{w_{\rm p}^2} = \frac{1}{w_{\rm p}'^2} + \frac{2}{w_{\rm SMF}^2} ,
\label{pompwyd}
\end{equation}
where $w_{\rm SMF}$ is the radius of the SMF Gaussian functions. Henceforth, $w_{\rm p}$ represents the effective pump mode size, instead of the original pump mode size, as before.

\subsection{Output (LG) modal basis}

The basis for the output density matrix is chosen to be the LG modes. In terms of normalized coordinates, these modes are given by
\begin{eqnarray}
E_{p,\ell}(u,v,t) & = & {\cal N} \frac{(u \pm iv)^{|\ell|} (1+it)^p}{(1-it)^{p+|\ell|+1}} \exp \left( \frac{u^2+v^2}{it-1} \right) \nonumber \\
& & \times L_p^{|\ell|} \left[ \frac{2(u^2+v^2)}{1+t^2} \right] ,
\label{lgm}
\end{eqnarray}
where $p$ and $\ell$ are the radial index and the azimuthal index (the $\pm$ sign is given by the sign of $\ell$), respectively, $L_p^{|\ell|}(\cdot)$ represents the associate Laguerre polynomials and ${\cal N}$ is a normalization constant given by
\begin{equation}
{\cal N} = \left[ \frac{2^{|\ell|+1}p!}{\pi (p+|\ell|)!} \right]^{1/2} .
\label{lgn}
\end{equation} 
The normalized coordinates are given by $u=x/w_0$, $v=y/w_0$ and $t=z/z_R=z\lambda/\pi w_0^2$, in terms of the waist radius $w_0$ and the Rayleigh range $z_R$ of the output basis.

The LG modes are OAM eigenstates \cite{allen}. This implies that an LG beam as a whole (and every photon in it) has a well-defined OAM. The amount of OAM in the beam is proportional to the azimuthal index $\ell$. OAM is conserved in the SPDC process \cite{arnaut,franke,mair}. This means that the sum of OAM of the signal and the idler photons equals the OAM of the pump photon. As a result, the SPDC state is entangled in terms of OAM. Hence, without turbulence, the output density matrix in the LG basis would be that of a highly entangled biphoton state.

We only use LG modes with $p=0$ for the output basis. Since we use the SPS approximation, we also set $z=t=0$. The output space is restricted to a three-dimensional basis for each photon. The basis elements are chosen symmetrically to be $\{E_{0,{-\ell}},E_{0,0},E_{0,\ell}\}$. For our experiment, we consider the three cases, $\ell=1,2,3$.

To expedite the calculations, we use a generating function for the LG modes \cite{ipe,pindex,noncol}. For $p=0$ and $z=t=0$, the generating function is given by
\begin{equation}
{\cal G}_{\pm} = \frac{1}{w_0} \exp \left[ \frac{(x\pm i y) \mu}{w_0} - \frac{x^2+y^2}{w_0^2} \right] ,
\label{genlg}
\end{equation} 
where $\mu$ is the generating parameter for the azimuthal index and the $\pm$ sign in the exponent is determined by the sign of $\ell$. To generate a particular mode, one uses
\begin{equation}
E_{0,\ell}({\bf x}) =  {\cal N} \left. \partial_{\mu}^{|\ell|} {\cal G}_{\pm} \right|_{\mu=0} ,
\label{simodes}
\end{equation}
where ${\cal N}$ is given in Eq.~(\ref{lgn}) with $p=0$.

Without turbulence, the (pure) state that one would obtain in the output within the ($3\times 3$)-dimensional space in which we measure, can be expressed as
\begin{equation}
\ket{\Psi} =  {\cal A}_0 \ket{0}_A \ket{0}_B + {\cal A}_{|\ell|} \left( \ket{\ell}_A \ket{{-\ell}}_B + \ket{{-\ell}}_A \ket{\ell}_B \right) ,
\label{suiwer}
\end{equation}
where the coefficients ${\cal A}_0$ and ${\cal A}_{|\ell|}$ are determined by the OAM spectrum of the SPDC state. For a maximally entangled state ${\cal A}_0={\cal A}_{|\ell|}=1/\sqrt{3}$. However, for the SPDC state we generally have ${\cal A}_0>{\cal A}_{|\ell|}$, which implies that the initial state is less than maximally entangled. Nevertheless, with appropriate experimental conditions, the initial entanglement of the state can be close to being maximally entangled.

\subsection{Generating function for density matrix elements}

Substitute, Eqs.~(\ref{struktq}), (\ref{spdcinx}) and (\ref{genlg}) into Eq.~(\ref{spsint}), and evaluate the eight integrals. One then imposes the thin crystal limit by setting $\beta=0$. The result is a generating function for the elements of the density matrix, given by 
\begin{eqnarray}
{\cal G}_{\rho} & = & M_0 M_1 \exp\left[ (M_0-M_1) \left(S^{(+)}_{13}\mu_1\mu_3 \right. \right. \nonumber \\
& & \left. + S^{(+)}_{14}\mu_1\mu_4 + S^{(+)}_{23}\mu_2\mu_3 + S^{(+)}_{24}\mu_2\mu_4\right) \nonumber \\
& & \left. +(M_0+M_1) \left(S^{(-)}_{12}\mu_1\mu_2 + S^{(-)}_{34}\mu_3\mu_4\right)\right] ,
\label{elmgen}
\end{eqnarray}
where we neglect an overall constant, $\mu_1 ... \mu_4$ are four generating parameters, respectively associated with the four indices $m,n,p,q$ in Eq.~(\ref{spsint}) and where
\begin{eqnarray}
M_0 & = & \frac{1}{2(\alpha+2)} \\
M_1 & = & \frac{1}{2(\alpha+2+\alpha\xi)} \\
S^{(\pm)}_{mn} & = & \left\{\begin{array}{cl} 1 & {\rm for} ~~ {\rm sign}(\ell_m) = \pm {\rm sign}(\ell_n) \\ 0 & {\rm otherwise}  \\ \end{array} \right. ,
\end{eqnarray}
with 
\begin{eqnarray}
\alpha & = & \frac{w_0^2}{w_{\rm p}^2} , \label{alphadef} \\
\xi & = & 6.88 \left(\frac{w_{\rm p}}{r_0}\right)^{5/3} .
\end{eqnarray}
To generate a particular matrix element, using Eq.~(\ref{elmgen}), one needs to compute Eq.~(\ref{simodes}) for each of the four $\ell$-values that are associated with $m,n,p,q$.

\subsection{Negativity}

There is currently no known entanglement measure for high dimensional states that is easy to compute and gives the exact amount of entanglement in the state. The concurrence \cite{wootters}, which is relatively easy to calculate for qubits, is much more complicated to compute when it is generalized to arbitrary dimensions. Nevertheless, one needs to quantify the high dimensional entanglement in order to understand the effects of turbulence on high dimensional OAM entangled states. Such an understanding is necessary to enable the successful implementation of a free-space quantum communication channel with high dimensional OAM states. In our experiment, we use the negativity to quantify the entanglement.

The negativity is defined as 
\begin{equation}
{\cal E} = \frac{1}{2} \sum_n (|\lambda_n|-\lambda_n) ,
\end{equation}
where $\lambda_n$ denotes the eigenvalues of the partial transpose of the density matrix. To compute the partial transpose of a bipartite density matrix, one exchanges the rows and columns for one of the two partites, while leaving those of the other one unchanged. The expressions for the negativity that we obtained from our theoretical calculations for $\ell=1,2,3$, are provided in Appendix \ref{negexp}. The curves of the negativity obtained from these theoretical calculations are presented together with the experimental results in Sec.~\ref{resdis}.

\section{Experimental setup}
\label{opstel}

\begin{figure}[th]
\includegraphics{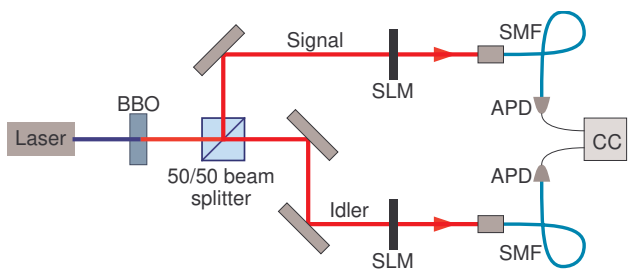}
\caption{Experimental setup used to perform high dimensional quantum state tomography on the quantum state after passing through SPS turbulence.}
\label{setup}
\end{figure}

The experimental setup is shown in Fig.~\ref{setup}. A mode-locked laser source with a wavelength of 355~nm, an average power of 350~mW and a repetition rate of 80~MHz pumps a 3~mm-thick type I BBO crystal to produce noncollinear, degenerate photon pairs via SPDC. A small noncollinear angle ($\sim 3$ degrees between signal and idler beams) is used to improve the OAM bandwidth \cite{romero,noncol}. The pump beam has a radius of 0.24~mm at the crystal. The plane of the crystal is imaged onto SLMs in the signal and idler beams, respectively, with a magnification of $\times 4$ (4-f system with $f_{1} = 100$~mm and $f_{2} = 400$~mm not shown). The SLM planes are re-imaged with a demagnification factor of $\times 375$ (4-f system with $f_{3} = 750$~mm and $\textrm{f}_{4} = 2$~mm not shown) onto single-mode fibres (SMFs). The radii of backprojected beams emitted from the SMFs onto the crystal plane are $\sim 0.26$~mm, giving an effective pump mode size of 0.15~mm. The down-converted light beams pass through 10~nm bandwidth interference filters (IF) before coupling into the SMFs. Avalanche photo diodes (APDs) at the ends of the SMFs are used to register the photon pairs with the aid of a coincidence counter (CC). The measured coincidence counts are accumulated over a 2~s integration time, with a gating time of 12.5~ns (based on the repetition rate).

Projective measurements are performed with the aid of the SLMs, by selecting particular pairs of LG modes (and superpositions of LG modes) for detection. We employ techniques for complex amplitude modulation \cite{arrizon1} on the phase-only SLMs to ensure that the modulation involves the exact LG mode functions and not only their phase functions. The size of the modes on the SLM is 0.45~mm, which gives $\alpha=0.59$.

The atmospheric turbulence is simulated in the experiment by adding a random phase fluctuation to the encoded modal basis functions on one of the SLMs. This random phase function is computed with \cite{knepp,mf1,dainty}
\begin{equation}
\theta({\bf x}) = \frac{1}{\Delta} {\cal F}^{-1} \left\{\chi({\bf a}) \left[\Phi_{\theta}({\bf a}) \right]^{1/2} \right\} ,
\end{equation}
where ${\cal F}^{-1}\{\cdot\}$ is the two-dimensional inverse Fourier transform, $\Delta$ is the sample spacing in the frequency domain and $\chi({\bf a})$ is a frequency domain delta-correlated zero-mean Gaussian random complex function. If we assume that $\theta({\bf x})$ is real-valued, then $\chi^*({\bf a})=\chi(-{\bf a})$. However, by allowing $\theta({\bf x})$ to be complex, one obtains two phase functions --- the real and the imaginary parts of $\theta({\bf x})$ --- with each calculation.

The phase power spectral density $\Phi_{\theta}$ is related to the refractive index power spectral density $\Phi_n$ through
\begin{equation}
\Phi_\theta({\bf a}) = 2\pi k^2 z \Phi_n(2\pi{\bf a},0) ,
\end{equation}
where $k=2\pi/\lambda$ is the wavenumber (not to be confused with $|{\bf k}|$ below). We use the refractive index power spectral density in Kolmogorov theory, given by \cite{kolmog,scintbook}
\begin{equation}
\Phi_n({\bf k}) = 0.033~C_n^2 |{\bf k}|^{-11/3} ,
\end{equation}
to calculate the phase screen. Subgrid sample points are added to the Fourier domain representation of the phase function to ensure that the calculated random phase functions can reproduce the Kolmogorov structure function reliably \cite{dainty}.

\section{Results and discussion}
\label{resdis}

We considered values of $W=w_0/r_0$, representing the scintillation strength of the random phase function, in the range 0 to 1.5. For each value of $W$, we computed 25 different sets of phase functions, corresponding to different realizations of the simulated turbulent medium. A quantum state tomography \cite{thew} is performed for each realization, to reconstruct the bipartite qutrit density matrix. The final density matrix representing the state of the two photons is calculated by averaging density matrices corresponding to each value of $W$. From the averaged density matrices, we then compute the negativity.

\begin{figure}[ht]
\includegraphics{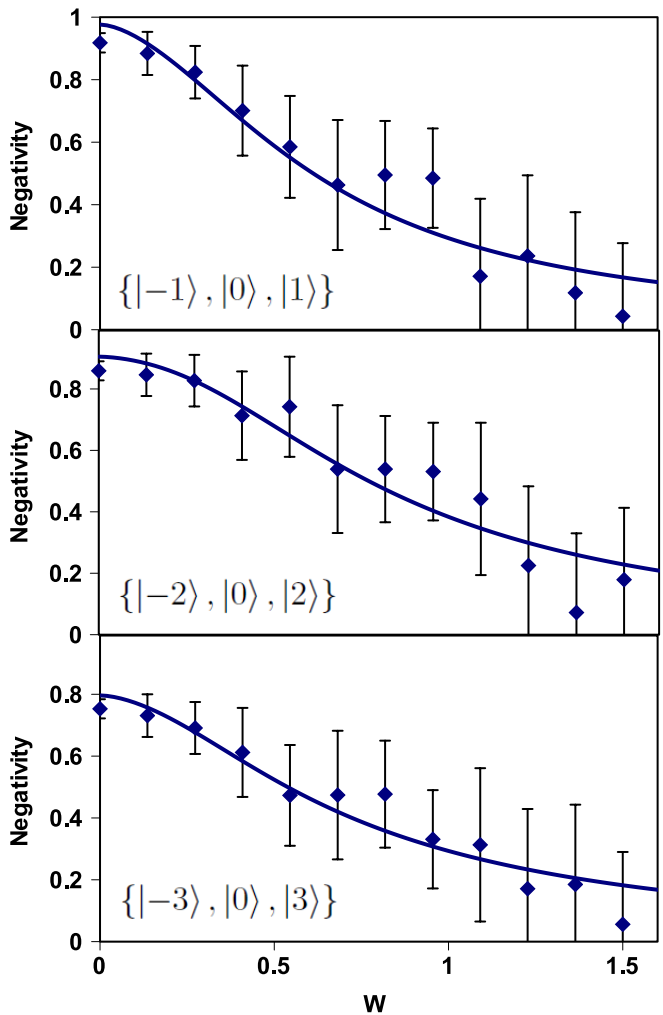}
\caption{The negativity is shown as a function of $W$ for $\ell=1,2,3$, respectively. The solid lines are theoretical curves and the points are experimental results. The error bars indicate the statistical error due to the Poisson photon statistics}
\label{negqut}
\end{figure}

The results are shown in Fig.~\ref{negqut}. One can see that the negativity decreases gradually with increasing $W$. When only one of the two photons propagates through turbulence, the theoretical value of the negativity never reaches zero. We see that the experimentally obtained negativities follow the same trend, at least up to $W=1.5$. The initial values for the negativity in the three graphs decrease gradually as the value of $|\ell|$ increases. This is an indication of the OAM spectrum that is produced in the SPDC process and measured in terms of the chosen measurement basis. It also depends on the value of $\alpha$.

\begin{figure}[ht]
\includegraphics{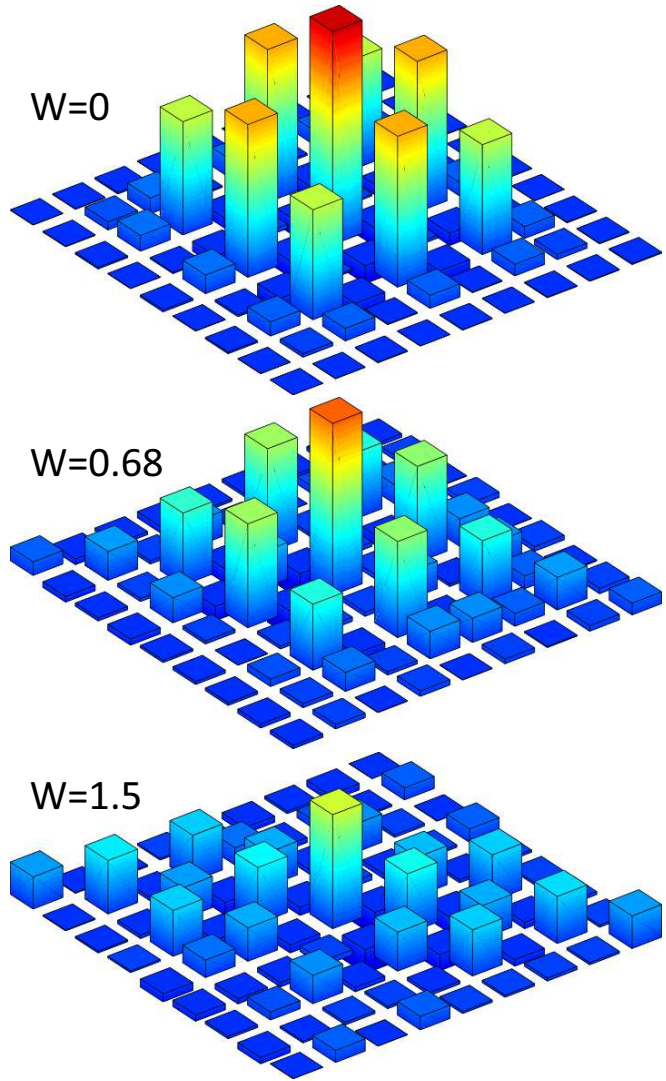}
\caption{Graphic representations of the density matrices for $W=0, 0.68, 1.5$, respectively. The diagonal elements of the densty matrices are horizontally arranged. }
\label{digt}
\end{figure}

For the case when $\ell=1$, we produced graphic representations of the three averaged density matrices, for which $W=0, 0.68, 1.5$, respectively. These are shown in Fig.~\ref{digt}. The density matrices for pairs of entangled qutrits are $9\times 9$ matrices. The magnitudes of the central (highest) elements are $0.45$, $0.36$ and $0.24$ for the three respective cases shown in Fig.~\ref{digt}. In the case where $W=0$, one can see the input state having a high purity. The varying heights (magnitudes of the elements) indicate that the input state is not maximally entangled, due to the modal spectrum that is produced in the SPDC process. For $W=0.68$, we see that other elements in the density matrix start to grow at the cost of the elements representing the original input state. Finally, for $W=1.5$, one observes that the other elements in the matrix, in particular those on the diagonal, start to dominate over the elements of the original input state.

\begin{figure}[ht]
\includegraphics{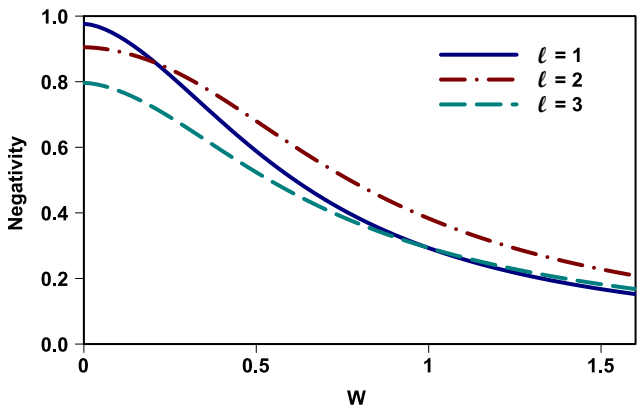}
\caption{Comparison of the theoretical negativity curves as a function of $W$.}
\label{thneg}
\end{figure}

The three theoretical curves from the three graphs in Fig.~\ref{negqut}, are shown together in Fig.~\ref{thneg}. One observes that, contrary to the case with qubits \cite{sr}, a higher value of $|\ell|$ does not give a better performance during propagation through turbulence. In fact, the curves cross each other at particular values of $W$. 

The original trend where higher value of $|\ell|$ performed better in turbulence, applied to qubits (Bell-states) within the SPS approximation (weak scintillation). It has already been shown previously that this trend does not apply in strong scintillation \cite{turbsim}. Here, it is shown that the trend also does not apply for higher dimensional states in weak scintillation. As a result, one would not expect to see the trend for higher dimensional states in strong scintillation.

\section{Conclusion}
\label{concl}

A quantum state that is produced by type I SPDC is entangled in its spatial degrees of freedom. This entanglement manifests strongly in the OAM degrees of freedom, because the Schmidt basis of this state is an OAM basis. Here, we study the evolution of this OAM entangled state through turbulence, within the weak scintillation conditions appropriate for a SPS approximation. The study includes both theoretical predictions, based on calculations using the SPS model, and experimental work, using an SLM to modulate one of the two entangled photons with a random phase function to simulate the turbulence.

Our experimental results for the evolution of high dimensional quantum states in an OAM basis show that the amount of entanglement between two qutrits does not decay slower when higher $|\ell|$-values are considered. This is in contrast to the situation that was found for qubits in weak scintillation, namely that the entanglement between two qubits evolving in turbulence is more robust when higher $|\ell|$-values are considered.

Comparing the results that we obtain here for the qutrits, with those that were previously obtained for the qubits \cite{oamturb} under the same conditions, we see that the entanglement for the qutrits decay quicker as a function of $W$ than the entanglement for the qubits. The benefit of using higher dimensional states (better security and higher information capacity), may therefore be offset by a poorer performance in turbulence. This would play an important role in the design of a free-space QKD system.

\vskip 10 mm

\section*{Acknowledgements}

This work was done with funding support from the CSIR and NRF.

\appendix
\section{Expressions for the negativity}
\label{negexp}

The expressions for the negativity obtained for $\ell=1,2,3$, all have the form
\begin{equation}
{\cal E}_{\ell} = \frac{4 (1+\eta) A_{\ell}}{B_{\ell}} ,
\end{equation}
where
\begin{equation}
\eta = \frac{6.88 \alpha W^{5/3}}{2+\alpha} .
\end{equation}
For the respective values of $\ell$, we have
\begin{widetext}
\begin{eqnarray}
A_1 & = & 3+\alpha \nonumber \\
B_1 & = & (1+\eta)^2\alpha^2+4(1+\eta)(1+2\eta)\alpha+4(3+6\eta+5\eta^2) \nonumber \\
A_2 & = & 2(6+10\eta+7\eta^2)+2(1+\eta)^2(4+\alpha)\alpha \nonumber \\
B_2 & = & 8(6+16\eta+24\eta^2+18\eta^3+7\eta^4)+4(1+\eta)^2[4(2+4\eta+3\eta^2)+(6+12\eta+7\eta^2)\alpha]\alpha+(1+\eta)^4(8+\alpha)\alpha^3 \nonumber \\
A_3 & = & 6(8+24\eta+36\eta^2+24\eta^3+9\eta^4)+(1+\eta)^2(4+6\eta+3\eta^2)(12+6\alpha+\alpha^2)\alpha \nonumber \\
B_3 & = & 16(12+48\eta+108\eta^2+138\eta^3+105\eta^4+45\eta^5+11\eta^6)+4(1+\eta)^3[12(4+12\eta+12\eta^2+5\eta^3) \nonumber \\ 
& & +6(10+30\eta+30\eta^2+11\eta^3)\alpha+(40+120\eta+120\eta^2+41\eta^3)\alpha^2]\alpha+(1+\eta)^6(60+12\alpha+\alpha^2)\alpha^4 .
\end{eqnarray}
\end{widetext}


\begin{thebibliography}{40}
\expandafter\ifx\csname natexlab\endcsname\relax\def\natexlab#1{#1}\fi
\expandafter\ifx\csname bibnamefont\endcsname\relax
  \def\bibnamefont#1{#1}\fi
\expandafter\ifx\csname bibfnamefont\endcsname\relax
  \def\bibfnamefont#1{#1}\fi
\expandafter\ifx\csname citenamefont\endcsname\relax
  \def\citenamefont#1{#1}\fi
\expandafter\ifx\csname url\endcsname\relax
  \def\url#1{\texttt{#1}}\fi
\expandafter\ifx\csname urlprefix\endcsname\relax\def\urlprefix{URL }\fi
\providecommand{\bibinfo}[2]{#2}
\providecommand{\eprint}[2][]{\url{#2}}

\bibitem[{\citenamefont{Walborn et~al.}(2006)\citenamefont{Walborn, Lemelle,
  Almeida, and Ribeiro}}]{walborn}
\bibinfo{author}{\bibfnamefont{S.~P.} \bibnamefont{Walborn}},
  \bibinfo{author}{\bibfnamefont{D.~S.} \bibnamefont{Lemelle}},
  \bibinfo{author}{\bibfnamefont{M.~P.} \bibnamefont{Almeida}},
  \bibnamefont{and} \bibinfo{author}{\bibfnamefont{P.~H.~S.}
  \bibnamefont{Ribeiro}}, \bibinfo{journal}{Phys. Rev. Lett.}
  \textbf{\bibinfo{volume}{96}}, \bibinfo{pages}{090501}
  (\bibinfo{year}{2006}).

\bibitem[{\citenamefont{Bechmann-Pasquinucci and Peres}(2000)}]{bp2000}
\bibinfo{author}{\bibfnamefont{H.}~\bibnamefont{Bechmann-Pasquinucci}}
  \bibnamefont{and} \bibinfo{author}{\bibfnamefont{A.}~\bibnamefont{Peres}},
  \bibinfo{journal}{Phys. Rev. Lett.} \textbf{\bibinfo{volume}{85}},
  \bibinfo{pages}{3313} (\bibinfo{year}{2000}).

\bibitem[{\citenamefont{Konrad et~al.}(2007)\citenamefont{Konrad, de~Melo,
  Tiersch, Kasztelan, Aragao, and Buchleitner}}]{konrad}
\bibinfo{author}{\bibfnamefont{T.}~\bibnamefont{Konrad}},
  \bibinfo{author}{\bibfnamefont{F.}~\bibnamefont{de~Melo}},
  \bibinfo{author}{\bibfnamefont{M.}~\bibnamefont{Tiersch}},
  \bibinfo{author}{\bibfnamefont{C.}~\bibnamefont{Kasztelan}},
  \bibinfo{author}{\bibfnamefont{A.}~\bibnamefont{Aragao}}, \bibnamefont{and}
  \bibinfo{author}{\bibfnamefont{A.}~\bibnamefont{Buchleitner}},
  \bibinfo{journal}{Nat. Phys.} \textbf{\bibinfo{volume}{4}},
  \bibinfo{pages}{99} (\bibinfo{year}{2007}).

\bibitem[{\citenamefont{Tiersch et~al.}(2008)\citenamefont{Tiersch, de~Melo,
  and Buchleitner}}]{tiersch}
\bibinfo{author}{\bibfnamefont{M.}~\bibnamefont{Tiersch}},
  \bibinfo{author}{\bibfnamefont{F.}~\bibnamefont{de~Melo}}, \bibnamefont{and}
  \bibinfo{author}{\bibfnamefont{A.}~\bibnamefont{Buchleitner}},
  \bibinfo{journal}{Phys. Rev. Lett.} \textbf{\bibinfo{volume}{101}},
  \bibinfo{pages}{170502} (\bibinfo{year}{2008}).

\bibitem[{\citenamefont{Gr{\"o}blacher
  et~al.}(2006)\citenamefont{Gr{\"o}blacher, Jennewein, Vaziri, Weihs, and
  Zeilinger}}]{zeil2006}
\bibinfo{author}{\bibfnamefont{S.}~\bibnamefont{Gr{\"o}blacher}},
  \bibinfo{author}{\bibfnamefont{T.}~\bibnamefont{Jennewein}},
  \bibinfo{author}{\bibfnamefont{A.}~\bibnamefont{Vaziri}},
  \bibinfo{author}{\bibfnamefont{G.}~\bibnamefont{Weihs}}, \bibnamefont{and}
  \bibinfo{author}{\bibfnamefont{A.}~\bibnamefont{Zeilinger}},
  \bibinfo{journal}{New J. of Phys.} \textbf{\bibinfo{volume}{8}},
  \bibinfo{pages}{75} (\bibinfo{year}{2006}).

\bibitem[{\citenamefont{Mafu et~al.}(2013)\citenamefont{Mafu, Dudley, Goyal,
  Giovannini, McLaren, Padgett, Konrad, Petruccione, L\"utkenhaus, and
  Forbes}}]{mafu}
\bibinfo{author}{\bibfnamefont{M.}~\bibnamefont{Mafu}},
  \bibinfo{author}{\bibfnamefont{A.}~\bibnamefont{Dudley}},
  \bibinfo{author}{\bibfnamefont{S.}~\bibnamefont{Goyal}},
  \bibinfo{author}{\bibfnamefont{D.}~\bibnamefont{Giovannini}},
  \bibinfo{author}{\bibfnamefont{M.}~\bibnamefont{McLaren}},
  \bibinfo{author}{\bibfnamefont{M.~J.} \bibnamefont{Padgett}},
  \bibinfo{author}{\bibfnamefont{T.}~\bibnamefont{Konrad}},
  \bibinfo{author}{\bibfnamefont{F.}~\bibnamefont{Petruccione}},
  \bibinfo{author}{\bibfnamefont{N.}~\bibnamefont{L\"utkenhaus}},
  \bibnamefont{and} \bibinfo{author}{\bibfnamefont{A.}~\bibnamefont{Forbes}},
  \bibinfo{journal}{Phys. Rev. A} \textbf{\bibinfo{volume}{88}},
  \bibinfo{pages}{032305} (\bibinfo{year}{2013}).

\bibitem[{\citenamefont{Capraro et~al.}(2012)\citenamefont{Capraro, Tomaello,
  Dall'{A}rche, Gerlin, Ursin, Vallone, and Villoresi}}]{capraro}
\bibinfo{author}{\bibfnamefont{I.}~\bibnamefont{Capraro}},
  \bibinfo{author}{\bibfnamefont{A.}~\bibnamefont{Tomaello}},
  \bibinfo{author}{\bibfnamefont{A.}~\bibnamefont{Dall'{A}rche}},
  \bibinfo{author}{\bibfnamefont{F.}~\bibnamefont{Gerlin}},
  \bibinfo{author}{\bibfnamefont{R.}~\bibnamefont{Ursin}},
  \bibinfo{author}{\bibfnamefont{G.}~\bibnamefont{Vallone}}, \bibnamefont{and}
  \bibinfo{author}{\bibfnamefont{P.}~\bibnamefont{Villoresi}},
  \bibinfo{journal}{Phys. Rev. Lett.} \textbf{\bibinfo{volume}{109}},
  \bibinfo{pages}{200502} (\bibinfo{year}{2012}).

\bibitem[{\citenamefont{Arnaut and Barbosa}(2000)}]{arnaut}
\bibinfo{author}{\bibfnamefont{H.~H.} \bibnamefont{Arnaut}} \bibnamefont{and}
  \bibinfo{author}{\bibfnamefont{G.~A.} \bibnamefont{Barbosa}},
  \bibinfo{journal}{Phys. Rev. Lett.} \textbf{\bibinfo{volume}{85}},
  \bibinfo{pages}{286} (\bibinfo{year}{2000}).

\bibitem[{\citenamefont{Franke-Arnold et~al.}(2002)\citenamefont{Franke-Arnold,
  Barnett, Padgett, and Allen}}]{franke}
\bibinfo{author}{\bibfnamefont{S.}~\bibnamefont{Franke-Arnold}},
  \bibinfo{author}{\bibfnamefont{S.~M.} \bibnamefont{Barnett}},
  \bibinfo{author}{\bibfnamefont{M.~J.} \bibnamefont{Padgett}},
  \bibnamefont{and} \bibinfo{author}{\bibfnamefont{L.}~\bibnamefont{Allen}},
  \bibinfo{journal}{Phys. Rev. A} \textbf{\bibinfo{volume}{65}},
  \bibinfo{pages}{033823} (\bibinfo{year}{2002}).

\bibitem[{\citenamefont{Mair et~al.}(2001)\citenamefont{Mair, Vaziri, Weihs,
  and Zeilinger}}]{mair}
\bibinfo{author}{\bibfnamefont{A.}~\bibnamefont{Mair}},
  \bibinfo{author}{\bibfnamefont{A.}~\bibnamefont{Vaziri}},
  \bibinfo{author}{\bibfnamefont{G.}~\bibnamefont{Weihs}}, \bibnamefont{and}
  \bibinfo{author}{\bibfnamefont{A.}~\bibnamefont{Zeilinger}},
  \bibinfo{journal}{Nature} \textbf{\bibinfo{volume}{412}},
  \bibinfo{pages}{313} (\bibinfo{year}{2001}).

\bibitem[{\citenamefont{Smith and Raymer}(2006)}]{sr}
\bibinfo{author}{\bibfnamefont{B.~J.} \bibnamefont{Smith}} \bibnamefont{and}
  \bibinfo{author}{\bibfnamefont{M.~G.} \bibnamefont{Raymer}},
  \bibinfo{journal}{Phys. Rev. A} \textbf{\bibinfo{volume}{74}},
  \bibinfo{pages}{062104} (\bibinfo{year}{2006}).

\bibitem[{\citenamefont{Gopaul and Andrews}(2007)}]{qturb4}
\bibinfo{author}{\bibfnamefont{C.}~\bibnamefont{Gopaul}} \bibnamefont{and}
  \bibinfo{author}{\bibfnamefont{R.}~\bibnamefont{Andrews}},
  \bibinfo{journal}{New J. Phys.} \textbf{\bibinfo{volume}{9}},
  \bibinfo{pages}{94} (\bibinfo{year}{2007}).

\bibitem[{\citenamefont{Jha et~al.}(2010)\citenamefont{Jha, Tyler, and
  Boyd}}]{qturb3}
\bibinfo{author}{\bibfnamefont{A.~K.} \bibnamefont{Jha}},
  \bibinfo{author}{\bibfnamefont{G.~A.} \bibnamefont{Tyler}}, \bibnamefont{and}
  \bibinfo{author}{\bibfnamefont{R.~W.} \bibnamefont{Boyd}},
  \bibinfo{journal}{Phys. Rev. A} \textbf{\bibinfo{volume}{81}},
  \bibinfo{pages}{053832} (\bibinfo{year}{2010}).

\bibitem[{\citenamefont{Roux}(2011)}]{ipe}
\bibinfo{author}{\bibfnamefont{F.~S.} \bibnamefont{Roux}},
  \bibinfo{journal}{Phys. Rev. A} \textbf{\bibinfo{volume}{83}},
  \bibinfo{pages}{053822} (\bibinfo{year}{2011}).

\bibitem[{\citenamefont{Gonzalez~Alonso and Brun}(2013)}]{toddbrun}
\bibinfo{author}{\bibfnamefont{J.~R.} \bibnamefont{Gonzalez~Alonso}}
  \bibnamefont{and} \bibinfo{author}{\bibfnamefont{T.~A.} \bibnamefont{Brun}},
  \bibinfo{journal}{Phys. Rev. A} \textbf{\bibinfo{volume}{88}},
  \bibinfo{pages}{022326} (\bibinfo{year}{2013}).

\bibitem[{\citenamefont{Leonhard et~al.}(2015)\citenamefont{Leonhard,
  Shatokhin, and Buchleitner}}]{leonhard}
\bibinfo{author}{\bibfnamefont{N.~D.} \bibnamefont{Leonhard}},
  \bibinfo{author}{\bibfnamefont{V.~N.} \bibnamefont{Shatokhin}},
  \bibnamefont{and}
  \bibinfo{author}{\bibfnamefont{A.}~\bibnamefont{Buchleitner}},
  \bibinfo{journal}{Phys. Rev. A} \textbf{\bibinfo{volume}{91}},
  \bibinfo{pages}{012345} (\bibinfo{year}{2015}).

\bibitem[{\citenamefont{Pors et~al.}(2011)\citenamefont{Pors, Monken, Eliel,
  and Woerdman}}]{pors}
\bibinfo{author}{\bibfnamefont{B.-J.} \bibnamefont{Pors}},
  \bibinfo{author}{\bibfnamefont{C.~H.} \bibnamefont{Monken}},
  \bibinfo{author}{\bibfnamefont{E.~R.} \bibnamefont{Eliel}}, \bibnamefont{and}
  \bibinfo{author}{\bibfnamefont{J.~P.} \bibnamefont{Woerdman}},
  \bibinfo{journal}{Opt. Express} \textbf{\bibinfo{volume}{19}},
  \bibinfo{pages}{6671} (\bibinfo{year}{2011}).

\bibitem[{\citenamefont{Malik et~al.}(2012)\citenamefont{Malik, O'Sullivan,
  Rodenburg, Mirhosseini, Leach, Lavery, Padgett, and Boyd}}]{malik}
\bibinfo{author}{\bibfnamefont{M.}~\bibnamefont{Malik}},
  \bibinfo{author}{\bibfnamefont{M.}~\bibnamefont{O'Sullivan}},
  \bibinfo{author}{\bibfnamefont{B.}~\bibnamefont{Rodenburg}},
  \bibinfo{author}{\bibfnamefont{M.}~\bibnamefont{Mirhosseini}},
  \bibinfo{author}{\bibfnamefont{J.}~\bibnamefont{Leach}},
  \bibinfo{author}{\bibfnamefont{M.~P.~J.} \bibnamefont{Lavery}},
  \bibinfo{author}{\bibfnamefont{M.~J.} \bibnamefont{Padgett}},
  \bibnamefont{and} \bibinfo{author}{\bibfnamefont{R.~W.} \bibnamefont{Boyd}},
  \bibinfo{journal}{Opt. Express} \textbf{\bibinfo{volume}{20}},
  \bibinfo{pages}{13195} (\bibinfo{year}{2012}).

\bibitem[{\citenamefont{Hamadou~Ibrahim
  et~al.}(2013)\citenamefont{Hamadou~Ibrahim, Roux, McLaren, Konrad, and
  Forbes}}]{oamturb}
\bibinfo{author}{\bibfnamefont{A.}~\bibnamefont{Hamadou~Ibrahim}},
  \bibinfo{author}{\bibfnamefont{F.~S.} \bibnamefont{Roux}},
  \bibinfo{author}{\bibfnamefont{M.}~\bibnamefont{McLaren}},
  \bibinfo{author}{\bibfnamefont{T.}~\bibnamefont{Konrad}}, \bibnamefont{and}
  \bibinfo{author}{\bibfnamefont{A.}~\bibnamefont{Forbes}},
  \bibinfo{journal}{Phys. Rev. A} \textbf{\bibinfo{volume}{88}},
  \bibinfo{pages}{012312} (\bibinfo{year}{2013}).

\bibitem[{\citenamefont{Goyal et~al.}(2016)\citenamefont{Goyal,
  Hamadou~Ibrahim, Roux, Konrad, and Forbes}}]{qkdturb}
\bibinfo{author}{\bibfnamefont{S.~K.} \bibnamefont{Goyal}},
  \bibinfo{author}{\bibfnamefont{A.}~\bibnamefont{Hamadou~Ibrahim}},
  \bibinfo{author}{\bibfnamefont{F.~S.} \bibnamefont{Roux}},
  \bibinfo{author}{\bibfnamefont{T.}~\bibnamefont{Konrad}}, \bibnamefont{and}
  \bibinfo{author}{\bibfnamefont{A.}~\bibnamefont{Forbes}},
  \bibinfo{journal}{J. Opt.} \textbf{\bibinfo{volume}{18}},
  \bibinfo{pages}{064002} (\bibinfo{year}{2016}).

\bibitem[{\citenamefont{Br\"unner and Roux}(2013)}]{bruenner}
\bibinfo{author}{\bibfnamefont{T.}~\bibnamefont{Br\"unner}} \bibnamefont{and}
  \bibinfo{author}{\bibfnamefont{F.~S.} \bibnamefont{Roux}},
  \bibinfo{journal}{New J. Phys.} \textbf{\bibinfo{volume}{15}},
  \bibinfo{pages}{063005} (\bibinfo{year}{2013}).

\bibitem[{\citenamefont{Paterson}(2005)}]{paterson}
\bibinfo{author}{\bibfnamefont{C.}~\bibnamefont{Paterson}},
  \bibinfo{journal}{Phys. Rev. Lett.} \textbf{\bibinfo{volume}{94}},
  \bibinfo{pages}{153901} (\bibinfo{year}{2005}).

\bibitem[{\citenamefont{Hamadou~Ibrahim
  et~al.}(2014)\citenamefont{Hamadou~Ibrahim, Roux, and Konrad}}]{turbsim}
\bibinfo{author}{\bibfnamefont{A.}~\bibnamefont{Hamadou~Ibrahim}},
  \bibinfo{author}{\bibfnamefont{F.~S.} \bibnamefont{Roux}}, \bibnamefont{and}
  \bibinfo{author}{\bibfnamefont{T.}~\bibnamefont{Konrad}},
  \bibinfo{journal}{Phys. Rev. A} \textbf{\bibinfo{volume}{90}},
  \bibinfo{pages}{052115} (\bibinfo{year}{2014}).

\bibitem[{\citenamefont{Agnew et~al.}(2011)\citenamefont{Agnew, Leach, McLaren,
  Roux, and Boyd}}]{tomo}
\bibinfo{author}{\bibfnamefont{M.}~\bibnamefont{Agnew}},
  \bibinfo{author}{\bibfnamefont{J.}~\bibnamefont{Leach}},
  \bibinfo{author}{\bibfnamefont{M.}~\bibnamefont{McLaren}},
  \bibinfo{author}{\bibfnamefont{F.~S.} \bibnamefont{Roux}}, \bibnamefont{and}
  \bibinfo{author}{\bibfnamefont{R.~W.} \bibnamefont{Boyd}},
  \bibinfo{journal}{Phys. Rev. A} \textbf{\bibinfo{volume}{84}},
  \bibinfo{pages}{062101} (\bibinfo{year}{2011}).

\bibitem[{\citenamefont{Arrizon}(2003)}]{arrizon1}
\bibinfo{author}{\bibfnamefont{V.}~\bibnamefont{Arrizon}},
  \bibinfo{journal}{Opt. Lett.} \textbf{\bibinfo{volume}{28}},
  \bibinfo{pages}{2521} (\bibinfo{year}{2003}).

\bibitem[{\citenamefont{Roux}(2014)}]{lindb}
\bibinfo{author}{\bibfnamefont{F.~S.} \bibnamefont{Roux}}, \bibinfo{journal}{J.
  Phys. A: Math. Theor.} \textbf{\bibinfo{volume}{47}}, \bibinfo{pages}{195302}
  (\bibinfo{year}{2014}).

\bibitem[{\citenamefont{Andrews and Phillips}(1998)}]{scintbook}
\bibinfo{author}{\bibfnamefont{L.~C.} \bibnamefont{Andrews}} \bibnamefont{and}
  \bibinfo{author}{\bibfnamefont{R.~L.} \bibnamefont{Phillips}},
  \emph{\bibinfo{title}{Laser Beam Propagation Through Random Media}}
  (\bibinfo{publisher}{SPIE}, \bibinfo{address}{Washington},
  \bibinfo{year}{1998}).

\bibitem[{\citenamefont{Fried}(1966)}]{fried}
\bibinfo{author}{\bibfnamefont{D.~L.} \bibnamefont{Fried}},
  \bibinfo{journal}{J. Opt. Soc. Am.} \textbf{\bibinfo{volume}{56}},
  \bibinfo{pages}{1372} (\bibinfo{year}{1966}).

\bibitem[{\citenamefont{Leader}(1978)}]{leader}
\bibinfo{author}{\bibfnamefont{J.~C.} \bibnamefont{Leader}},
  \bibinfo{journal}{J. Opt. Soc. Am.} \textbf{\bibinfo{volume}{68}},
  \bibinfo{pages}{175} (\bibinfo{year}{1978}).

\bibitem[{\citenamefont{Law and Eberly}(2004)}]{eberly}
\bibinfo{author}{\bibfnamefont{C.~K.} \bibnamefont{Law}} \bibnamefont{and}
  \bibinfo{author}{\bibfnamefont{J.~H.} \bibnamefont{Eberly}},
  \bibinfo{journal}{Phys. Rev. Lett.} \textbf{\bibinfo{volume}{92}},
  \bibinfo{pages}{127903} (\bibinfo{year}{2004}).

\bibitem[{\citenamefont{Allen et~al.}(1992)\citenamefont{Allen, Beijersbergen,
  Spreeuw, and Woerdman}}]{allen}
\bibinfo{author}{\bibfnamefont{L.}~\bibnamefont{Allen}},
  \bibinfo{author}{\bibfnamefont{M.~W.} \bibnamefont{Beijersbergen}},
  \bibinfo{author}{\bibfnamefont{R.~J.~C.} \bibnamefont{Spreeuw}},
  \bibnamefont{and} \bibinfo{author}{\bibfnamefont{J.~P.}
  \bibnamefont{Woerdman}}, \bibinfo{journal}{Phys. Rev. A}
  \textbf{\bibinfo{volume}{45}}, \bibinfo{pages}{8185} (\bibinfo{year}{1992}).

\bibitem[{\citenamefont{Zhang et~al.}(2014)\citenamefont{Zhang, Roux, McLaren,
  and Forbes}}]{pindex}
\bibinfo{author}{\bibfnamefont{Y.}~\bibnamefont{Zhang}},
  \bibinfo{author}{\bibfnamefont{F.~S.} \bibnamefont{Roux}},
  \bibinfo{author}{\bibfnamefont{M.}~\bibnamefont{McLaren}}, \bibnamefont{and}
  \bibinfo{author}{\bibfnamefont{A.}~\bibnamefont{Forbes}},
  \bibinfo{journal}{Phys. Rev. A} \textbf{\bibinfo{volume}{89}},
  \bibinfo{pages}{043820} (\bibinfo{year}{2014}).

\bibitem[{\citenamefont{Zhang and Roux}(2014)}]{noncol}
\bibinfo{author}{\bibfnamefont{Y.}~\bibnamefont{Zhang}} \bibnamefont{and}
  \bibinfo{author}{\bibfnamefont{F.~S.} \bibnamefont{Roux}},
  \bibinfo{journal}{Phys. Rev. A} \textbf{\bibinfo{volume}{89}},
  \bibinfo{pages}{063802} (\bibinfo{year}{2014}).

\bibitem[{\citenamefont{Wootters}(1998)}]{wootters}
\bibinfo{author}{\bibfnamefont{W.~K.} \bibnamefont{Wootters}},
  \bibinfo{journal}{Phys. Rev. Lett.} \textbf{\bibinfo{volume}{80}},
  \bibinfo{pages}{2245} (\bibinfo{year}{1998}).

\bibitem[{\citenamefont{Romero et~al.}(2012)\citenamefont{Romero, Giovannini,
  Franke-Arnold, Barnett, and Padgett}}]{romero}
\bibinfo{author}{\bibfnamefont{J.}~\bibnamefont{Romero}},
  \bibinfo{author}{\bibfnamefont{D.}~\bibnamefont{Giovannini}},
  \bibinfo{author}{\bibfnamefont{S.}~\bibnamefont{Franke-Arnold}},
  \bibinfo{author}{\bibfnamefont{S.~M.} \bibnamefont{Barnett}},
  \bibnamefont{and} \bibinfo{author}{\bibfnamefont{M.~J.}
  \bibnamefont{Padgett}}, \bibinfo{journal}{Phys. Rev. A}
  \textbf{\bibinfo{volume}{86}}, \bibinfo{pages}{012334}
  (\bibinfo{year}{2012}).

\bibitem[{\citenamefont{Knepp}(1983)}]{knepp}
\bibinfo{author}{\bibfnamefont{D.~L.} \bibnamefont{Knepp}},
  \bibinfo{journal}{Proc. IEEE} \textbf{\bibinfo{volume}{71}},
  \bibinfo{pages}{722} (\bibinfo{year}{1983}).

\bibitem[{\citenamefont{Martin and Flatt{\'e}}(1988)}]{mf1}
\bibinfo{author}{\bibfnamefont{J.~M.} \bibnamefont{Martin}} \bibnamefont{and}
  \bibinfo{author}{\bibfnamefont{S.~M.} \bibnamefont{Flatt{\'e}}},
  \bibinfo{journal}{Appl. Opt.} \textbf{\bibinfo{volume}{27}},
  \bibinfo{pages}{2111} (\bibinfo{year}{1988}).

\bibitem[{\citenamefont{Lane et~al.}(1992)\citenamefont{Lane, Glindemann, and
  Dainty}}]{dainty}
\bibinfo{author}{\bibfnamefont{R.~G.} \bibnamefont{Lane}},
  \bibinfo{author}{\bibfnamefont{A.}~\bibnamefont{Glindemann}},
  \bibnamefont{and} \bibinfo{author}{\bibfnamefont{J.~C.}
  \bibnamefont{Dainty}}, \bibinfo{journal}{Waves in Random Media}
  \textbf{\bibinfo{volume}{2}}, \bibinfo{pages}{209} (\bibinfo{year}{1992}).

\bibitem[{\citenamefont{Kolmogorov}(1922)}]{kolmog}
\bibinfo{author}{\bibfnamefont{A.~N.} \bibnamefont{Kolmogorov}},
  \bibinfo{journal}{C. R. (Doki) Acad. Sci. U.S.S.R.}
  \textbf{\bibinfo{volume}{30}}, \bibinfo{pages}{301} (\bibinfo{year}{1922}).

\bibitem[{\citenamefont{Thew et~al.}(2002)\citenamefont{Thew, Nemoto, White,
  and Munro}}]{thew}
\bibinfo{author}{\bibfnamefont{R.~T.} \bibnamefont{Thew}},
  \bibinfo{author}{\bibfnamefont{K.}~\bibnamefont{Nemoto}},
  \bibinfo{author}{\bibfnamefont{A.~G.} \bibnamefont{White}}, \bibnamefont{and}
  \bibinfo{author}{\bibfnamefont{W.~J.} \bibnamefont{Munro}},
  \bibinfo{journal}{Phys. Rev. A} \textbf{\bibinfo{volume}{66}},
  \bibinfo{pages}{012303} (\bibinfo{year}{2002}).

\end{thebibliography}
\end{document}